\documentclass[twocolumn]{aastex631}

\def\simlt{\mathrel{\hbox{\rlap{\hbox{\lower4pt\hbox{$\sim$}}}\hbox{$<$}}}}
\def\simgt{\mathrel{\hbox{\rlap{\hbox{\lower4pt\hbox{$\sim$}}}\hbox{$>$}}}}
\def\s{\phantom{1}}

\def\simlt{\mathrel{\hbox{\rlap{\hbox{\lower4pt\hbox{$\sim$}}}\hbox{$<$}}}}
\def\simgt{\mathrel{\hbox{\rlap{\hbox{\lower4pt\hbox{$\sim$}}}\hbox{$>$}}}}

\def\chandra{Chandra}
\def\xmm{XMM-Newton}

\def\nicer{NICER}
\def\ksnr{Kes~79}
\def\kpsr{PSR~J1852$+$0040}
\def\esnr{PKS~1209$-$51/52}
\def\ppks{1E~1207.4$-$5209}
 
\def\ppsr{PSR~J0821$-$4300}
\def\psnr{Puppis~A}

\shorttitle{Noisy Timing Behavior of CCOs}
\shortauthors{Perez, Gotthelf \& Halpern}

\begin{document}

\title{Noisy Timing Behavior is a Feature of Central Compact Object Pulsars}

\author[0000-0002-6341-4548]{K.~I. Perez}
\affiliation{Department of Astronomy, Columbia University, 550 West 120th Street, New York, NY 10027, USA}

\author[0000-0003-3847-3957]{E.~V. Gotthelf}
\affiliation{Columbia Astrophysics Laboratory, Columbia University, 550 West 120th Street, New York, NY 10027, USA}

\author[0000-0003-4814-2377]{J.~P. Halpern}
\affiliation{Department of Astronomy, Columbia University, 550 West 120th Street, New York, NY 10027, USA}

\begin{abstract}
  We present a timing study of the three known central compact object (CCO) pulsars, isolated cooling neutron stars in supernova remnants (SNRs), using observations from \chandra, \xmm\ and \nicer\ spanning two decades. Relative to canonical young pulsars, CCOs are spinning down at a very slow rate $|\dot f| <10^{-15}$~s$^{-2}$, implying a surface dipole magnetic field strength $B_s < 10^{11}$~G that is too weak to account for their X-ray-emitting hot spots. Two CCO pulsars with sufficiently long monitoring, \ppks\ and \ppsr, are seen to deviate from steady spin-down; their timing residuals can be modeled by one or more glitches in $f$ and $\dot f$, or alternatively by extreme timing noise.  For the third CCO pulsar, \kpsr, the sparse temporal coverage was insufficient to detect such effects.  Glitch activity and timing noise in large samples of rotation-powered pulsars correlate best with $\dot f$, while the timing irregularities of the first two CCOs are extreme compared to pulsars of the same $\dot f$.  Nevertheless, timing activity in CCOs may arise from properties that they share with other young but more energetic pulsars: high internal temperature, strong buried magnetic field, and superfluid behavior.  Alternatively, continuing low-level accretion of supernova debris is not ruled out as a source of timing noise in CCOs.
  
\medskip\noindent
   {\it Unified Astronomy Thesaurus concepts:} Pulsars (1306); Neutron stars (1108); Compact objects (288); Pulsar timing method (1305)
\end{abstract}

\section {Introduction}

Central compact objects (CCOs), comprising a dozen or so cooling neutron stars (NSs) in supernova remnants (SNRs), are defined and distinguished from canonical, young rotation-powered pulsars by their steady surface thermal X-ray emission, lack of surrounding pulsar wind nebula, and nondetection at any other wavelength\footnote{ Recently, the first detection of weak radio pulsations from a CCO, \ppks, was reported \citep{zha25}.} (\citealt{pav02}; see \citealt{del17} for a review). Considering that their faintness and spectral softness are obstacles to widespread discovery in the Galactic plane, CCOs may be as common as more conspicuous NS classes, e.g., magnetars, and may represent a major channel of NS birth.

X-ray pulsations, evidently due to nonuniform surface temperature, have been detected from only three of the CCOs: \ppks\ in SNR \esnr, \ppsr\ in SNR \psnr\ and \kpsr\ in SNR \ksnr.  They have typical spin periods $P=(0.424, 0.112, 0.104$)~s but small spin-down rates $\dot f = (-1.2,-6.8,-7.9)\times 10^{-16}$~s$^{-2}$, implying surface dipole magnetic field strengths of $(9.8,2.9,3.1) \times 10^{10}$~G, respectively, which are exceptionally weak for such young pulsars \citep{got13,hal10}. 

The ages of the pulsars are the same as those of their SNRs, estimated as $t=(14,10,5)\times 10^3$~yr for \esnr\ \citep{epp24}, \psnr\ \citep{aru22} and \ksnr\ \citep{may21}, respectively. In contrast, when a pulsar ``characteristic age'' is quoted, defined as $\tau_c \equiv P/2\dot P$, the value is in the range $(2-3)\times 10^8$~yr for CCOs.  Because the relation between $\tau_c$ and true age $t$ is $t=\tau_c[1-(P_0/P)^2]$ for dipole braking with constant magnetic field, this spectacular difference simply indicates that the current spin period $P$ of the pulsar is likely identical to its otherwise unknown period at birth $P_0$ because its $B$ field is so weak.

Crucially, the surface magnetic field strength of \ppks\ inferred from spin-down, $B_s = 9.8\times10^{10}$~G, is close to $B\approx8\times10^{10}$~G, the value measured from its unique series of spectroscopic absorption features \citep{big03} interpreted as the electron cyclotron fundamental at 0.7~keV and its harmonics. That similar values of field strength are obtained by the two independent methods is persuasive evidence of the weak surface $B$ field for at least this CCO.

The remainder of the CCOs, having similar continuum spectral properties, have eluded searches for X-ray pulsations \citep{alf23}.  It is postulated that they may have even weaker magnetic fields, more uniform surface temperature distributions, or unfavorable geometry of hot spots. \citet{alf22} derived that the hot spots on the surface of \ppsr\ are probably close to its spin axis, and this could be the case for CCOs in general \citep{alf23}, which would limit the possible amplitude and detectability of pulsation. A similar conclusion about the magnetic axis of \ppks, that it is nearly aligned with the spin, was tentatively drawn from the polarization swing and width of its radio pulse \citep{zha25}.

It is even difficult to create the hot spots needed to produce X-ray pulsations in CCOs assuming only their weak surface dipoles.  The only mechanism thought to be capable of creating a nonuniform surface temperature is anisotropic heat conduction in a strong magnetic field.   As recounted in \citet{got18}, models that include strong nondipolar components, mainly toroidal ones in the crust, have long been studied.  More recently, \citet{igo21} calculated the magnetothermal evolution of strong crustal fields generated by a stochastic dynamo, finding a tangled geometry with negligible dipole component.  This creates factors of $\sim2$ difference in surface temperature on small scales, sufficient to produce the observed pulsations in CCOs.

Postulating a temporary suppression of the surface dipole field was the key to an explanation for the conspicuous absence of CCO descendants, the large population of weak $B$-field pulsars that should remain in the same location of the $P-\dot P$ diagram after their SNRs have faded.  If CCOs are actually born with a canonical NS magnetic dipole field that was promptly buried by fallback of a small amount $(\sim10^{-4}\,M_{\odot})$ of supernova debris, the buried field will diffuse back to the surface on a timescale of $\sim10^5$~yr \citep{mus95,ber10,ho11,vig12}.  A CCO will move vertically up in the $P-\dot P$ diagram as its dipole field grows back \citep{bog14,ho15,luo15}, merging with the population of ordinary radio pulsars.

The discovery of a glitch and/or strong timing noise from \ppks\ \citep{got18,got20} was of timely relevance to these theories of CCOs. Glitches are thought to result from ``starquakes'', stress relief of the NS crust, or from unpinning of superfluid vortices in the inner part of the NS crust \citep[e.g.,][]{lin98}. A review of glitch observations and theory is given by \citet{ant22}, while comprehensive compilations of glitch statistics can be found in \citet{esp11}, \citet{fue17} and \citet{bas22}. Timing noise has been attributed to variability in the interaction of the crustal superfluid with the Coulomb lattice of the solid crust \citep{jon90}, turbulence of the superfluid \citep{mel14}, or fluctuations in the structure of the magnetosphere, e.g., state switching \citep{lyn10}. Analysis methods for timing noise are reviewed in \citet{par19} and \citet{nam19}.

The timing irregularities in \ppks\ were extreme relative to the general pulsar population, in which glitch activity is mainly correlated with the frequency derivative $\dot f$.  Glitches are therefore unexpected in CCOs due to their small $\dot f$. However, the idea of stronger internal $B$ fields, needed to explain the CCOs' surface hot spots and population statistics, may find additional application in producing the timing irregularities of CCOs.  \cite{ho15} proposed that glitches could be triggered by the motion of strong magnetic fields through the NS crust, interacting with the neutron superfluid there, and that glitches would be a way of identifying CCO descendants.

To add complexity, it has not been ruled out that an entirely different process, low-level accretion in the propeller regime, could cause the observed timing noise, specifically the observed changes in $\dot f$ \citep{hal07,got18}.  This is because the required accretion rate of $\le10^{11}$~g~s$^{-1}$ makes a negligible contribution to the observed luminosity, while the mass supply could be a residual fallback disk of $\sim10^{-6}\,M_{\odot}$, much less mass than may have fallen back on the NS initially.  The spin parameters of CCOs fall into a regime where dipole braking and accretion torque noise could be comparable.

The subject of this paper is the long-term spin-down evolution of all three CCO pulsars spanning the two decades since their discovery. In Section~\ref{sec:timing} we present new observations from \chandra, \xmm, and \nicer\ that we use to update timing solutions, explore alternative timing models, and identify timing irregularities in \ppsr\ for the first time.  Section~\ref{sec:discussion} discusses the implications of the candidate glitches and timing noise and compares the CCO results with those of the general pulsar population.  Section~\ref{sec:conclusions} summarizes our conclusions.

\section {Timing Analysis}
\label{sec:timing}

We obtained new timing observations for the three CCO pulsars as part of our continued monitoring programs using the \chandra\ and \xmm\ observatories. We were able to achieve better temporal coverage by requesting observations from the two missions, allowing for their unique scheduling constraints. With the onset of the \nicer\ mission, we continued our coverage of \ppks\ using \nicer\ exclusively. The previously published datasets also included here are fully described in our previous work \citep[\ppks\ --- ][]{got07,hal11,got13,hal15,got18,got20}, \citep[\ppsr\ --- ][]{got09,got13}, and \citep[\kpsr\ ---][]{got05,hal07,hal10}. All of the observations used here are listed in the Appendix.

All datasets were reprocessed and reduced using the latest software for each mission, following the methods described in our previous papers. Photon arrival times were converted to barycentric dynamical time (TDB) using the DE405 solar system ephemeris and the \chandra\ coordinates given in the tables below for each object. For \ppsr, we account for its high proper motion using the ephemeris of \cite{may20}. 
Photons were extracted using a circular aperture optimized for each target and instrument, based on the local SNR contamination in the aperture.

For each observation, we generated a phase zero time-of-arrival (ToA) measurement to use in determining the long-term spin evolution of the NS, as follows. Starting from the filtered event files, we measured the pulse period at each epoch using a periodogram search for the maximum power around the expected frequency, as informed by our earlier work. For \ppks\ and \ppsr, their pulse profiles are well characterized by a sinusoidal function, and we computed phase zero from the Fourier coefficients of the unbinned photon arrival times. For consistency with \citet{got20}, we define phase zero as the minimum of the pulse for \ppks.  For \ppsr\ we use the maximum of the pulse as phase zero.  For \kpsr, which has a much broader pulse profile, a high-significance pulse profile template was generated by adding all the folded profiles, aligned by cross-correlating with an iterated template. We define phase zero for this CCO as the middle of the valley in the pulse profile.

Using the {\tt TEMPO} software \citep{nic15}, we initiated a phase-connected timing solution to a subset of ToAs for each pulsar, using a model for the rotation phase of the pulsar that includes one or two of its frequency derivatives: $$ \phi(t) = \phi_{o} + f(t-t_{o}) + {1\over2}\dot f (t-t_{o})^{2} + {1\over6}\ddot f (t-t_{o})^{3}.$$ Generally, we obtained optimally spaced sets of observations that allow us to start and maintain a phase-connected timing solution that accounts for every turn of the pulsar, with increased accuracy, by bootstrapping to longer intervals. This is usually possible unless the interval to the next ToA is too large to predict its cycle count by less than $\sim 10\%$ in phase. As described below, due to data gaps and timing irregularities, it was necessary at some point to restart a new solution for each pulsar to continue to follow and/or interpret its spin-down behavior.  However, if the timing residuals begin to deviate significantly and systematically from the trial solution, it signifies that the ephemeris no longer predicts the pulsar's spin-down, and more complex models need to be considered.

\subsection{\ppks\ in \esnr}
\label{sec:one}

In this work, we use all available data sets for \ppks\ from \xmm, \chandra, and \nicer, as listed in the Appendix. These span 23 yr and include 3.5 yr of monitoring since our previously published results \citep{got20}. The new observations used here were acquired exclusively using \nicer.  In the following, we extend the post-2015 glitch ephemeris and update the alternative models suggested by the earlier work. We do not repeat the previously reported fit for a possible pre-2015 glitch from \ppks, as that result is unchanged. In the following analysis, we limit the photon energy range to 0.5--1.6~keV, which is optimal for the pulsar's observed soft X-ray spectrum.

\begin{figure}
\centerline{\includegraphics[width=0.55\textwidth]{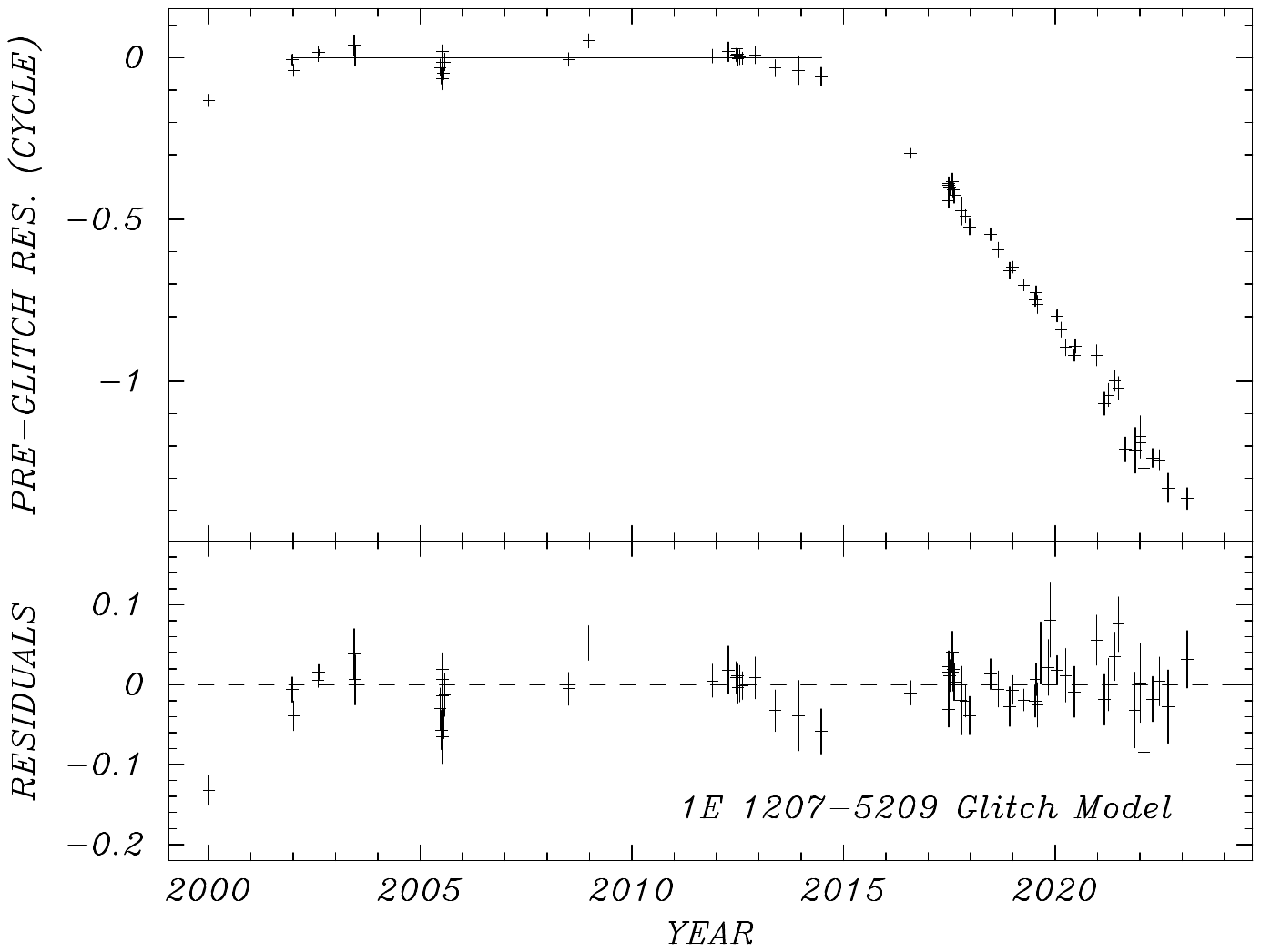}}
\caption{Glitch model for the spin-down of \ppks. Top: pulse-phase residuals from the pre-glitch  timing solution presented in Table~\ref{tab:pksglitch}.  The glitch epoch of 2015 January 19 is estimated by matching the frequency of the pre- and postglitch solutions. The year 2000 \chandra\ data point is not included in the fit (see Section~\ref{sec:one} for details). Bottom: combined residuals from fits to independent timing models for the pre- and postglitch intervals, respectively. The overall $\chi^2_{\nu}$ statistic for the fit is 1.49 for 60 DoF, taking into account the fit parameters for each interval. }
\label{fig:pksglitch}
\end{figure}

Figure~\ref{fig:pksglitch} graphs the ToA phase residuals from the pre-glitch timing solution given in Table~\ref{tab:pksglitch}, developed using data points from 2002 to 2014 \citep[as in][]{got20}.  As noted in our earlier work, the very first ToA, the \chandra\ observation of 2000, does not appear consistent with the preglitch timing solution and may indicate an even earlier glitch. We again excluded this data point from the fit but continue to plot it in Figure~\ref{fig:pksglitch} for reference.

After 2015, deviations from the expected phase of this solution were evident, and we were no longer able to extend the phase-connected solution forward. Instead, we postulated a glitch and found that the subsequent ToAs could be characterized by a new, postglitch phase-connected ephemeris. In this work, we have extended the postglitch ephemeris, which remains consistent with the earlier result and shows no evidence of a newer glitch. The updated change of frequency between the post- and preglitch timing solutions is $\Delta f = (3.87\pm0.39)\times10^{-9}$~Hz, with a glitch magnitude of $\Delta f/f_{\rm pred} = (1.64\pm0.16)\times10^{-9}$. The postglitch solution matches $f_{\rm pred}$, the frequency predicted by the preglitch solution on 2015 January~19 (MJD~57041), which we estimate as the epoch of the glitch (ignoring any short-time-scale recovery behavior, which would not have been sampled).  Of note, the glitch magnitude for \ppks\ is typical of rotation-powered pulsars with strong magnetic fields and rapid spin-down, including the Crab.  For reference, $\Delta f/f= 3\times 10^{-9}$ is where the lower-amplitude peak in the bimodal distribution of glitch magnitudes is centered \citep{esp11}.  The magnitude of the glitch is 23\% smaller than the value quoted in \citet{got20}.  Nevertheless, this is the same postglitch timing solution as the one published previously in the sense that the cycle counts calculated over the common time span are the same.  The fit statistic of the combined pre- and postglitch data sets is $\chi^2_{\nu} = 1.49$ for 60 DoF. This should be compared to $\chi^2_{\nu} = 1.44$ for 42 DoF previously reported for the shorter span. 

\begin{deluxetable}{lc}
\tighten
\tablewidth{0pt}
\tablecolumns{2}
\tablecaption{Glitch Ephemerides for \ppks}
\tablehead{
\colhead{Parameter} & \colhead{Value\tablenotemark{a} \hfill} 
}
\startdata
R.A. (J2000)                                  & $12^{\rm h}10^{\rm m}00^{\rm s}\!.91$ \\
Decl. (J2000)                                 & $-52^{\circ}26^{\prime}28^{\prime\prime}\!.4$ \\
Surface dipole magnetic field, $B_s$            & $9.8 \times 10^{10}$ G\\
Spin-down luminosity, $\dot E$                & $1.1 \times 10^{31}$ erg s$^{-1}$ \\
Characteristic age, $\tau_c\equiv P/2\dot P$  & 303~Myr \\
\cutinhead{Preglitch Timing Solution (2002--2014) \hfill}
 Epoch of ephemeris (MJD TDB)\tablenotemark{b}              &          54547.00000198           \\   
 Span of ephemeris (MJD)                   &      52266--56829                   \\
 Frequency, $f$                      &      2.357763492491(28) s$^{-1}$             \\
 Frequency derivative, $\dot f$  &     $-1.2317(66)\times 10^{-16}$ s$^{-2}$  \\
 Period, $P$                            &      0.4241307506816(50) s          \\
 Period derivative, $\dot P$               &      $2.216(12)\times 10^{-17}$     \\
 $\chi^2_{\nu}[{\rm DoF}]$                 &       1.80[25]                     \\
\cutinhead{Postglitch Timing Solution (2016--2023) \hfill}  
 Epoch of ephemeris (MJD TDB)\tablenotemark{b}              &          58144.00000219           \\
 Span of ephemeris (MJD)                   &      57597--59986                   \\
 Frequency, $f$                      &      2.35776345898(14)  s$^{-1}$           \\
 Frequency derivative, $\dot f$ &      $-1.120(27)\times 10^{-16}$ s$^{-2}$   \\
 Period, $P$                            &      0.424130756710(26) s           \\
 Period derivative, $\dot P$               &      $2.015(49)\times 10^{-17}$     \\
 $\chi^2_{\nu}[{\rm DoF}]$                 &       1.26[35]                     \\
\noalign{\vskip 0.5em}\hline\noalign{\vskip 0.5em}
 Glitch epoch (MJD)\tablenotemark{c}                        &      57041                    \\
 $\Delta f$                        &      $3.87(39)\times 10^{-9}$ s$^{-1}$ \\
 $\Delta f/f_{\rm pred}$                         &      $1.64(16)\times 10^{-9}$ \\
$\Delta \dot f$                        & $1.12(28)\times 10^{-17}$ s$^{-2}$ \\
 $\Delta \dot f/\dot f$                         &      $-0.091(23)$
\enddata
\tablenotetext{}{Notes. Derived parameters ($B_s$, $\dot E$, $\tau_c$) are based on the preglitch timing solution.}
\tablenotetext{a}{$1\sigma$ uncertainties in the last digits are given in parentheses.}  
\tablenotetext{b}{Epoch of the ephemeris corresponds to the minimum of the pulse.}
\tablenotetext{c}{Epoch of the glitch estimated by matching the frequency of the two timing solutions; this assumes a constant postglitch $\dot f$.}
\label{tab:pksglitch}
\end{deluxetable}

\begin{figure}
\centerline{\includegraphics[width=0.55\textwidth]{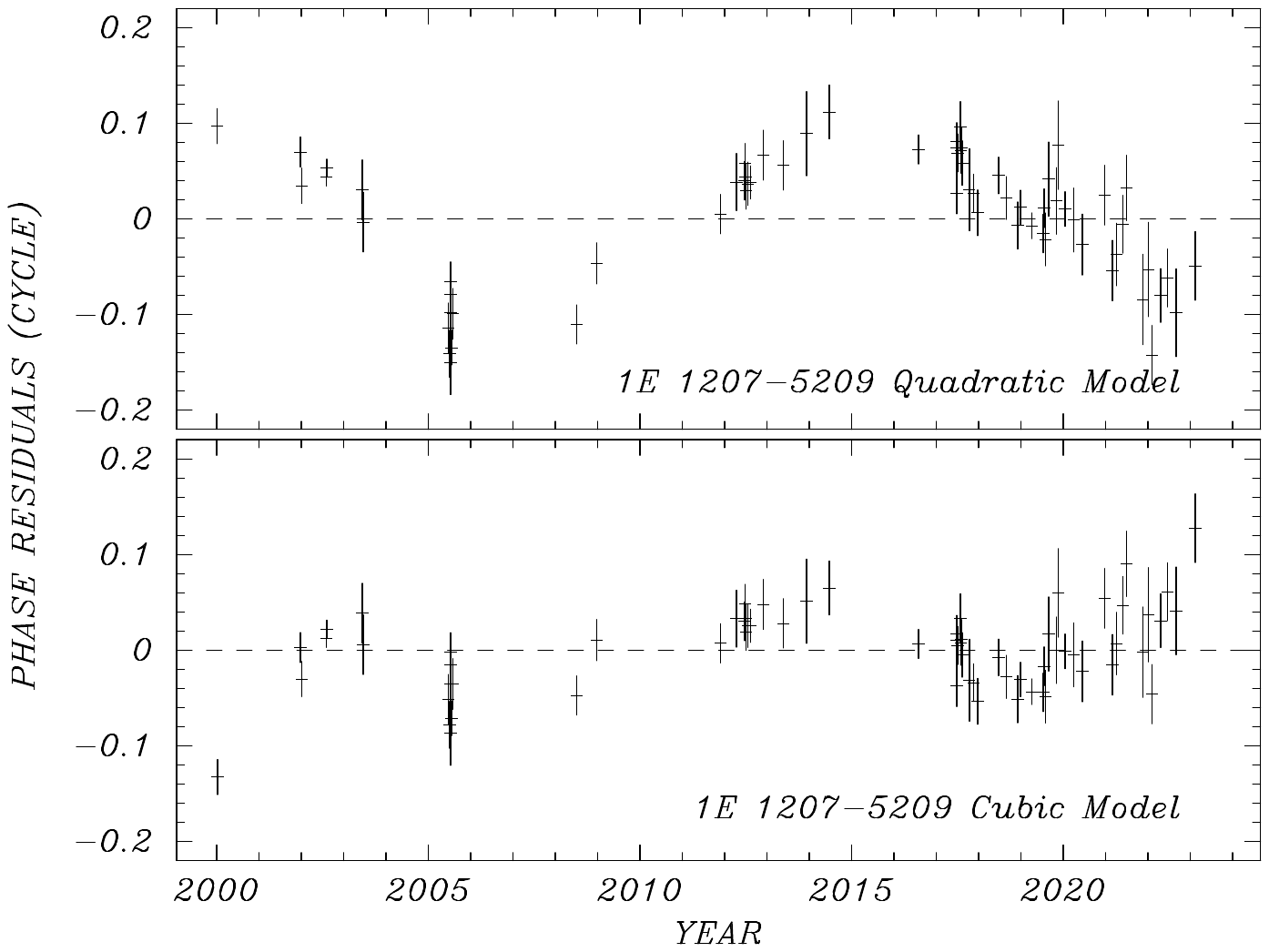}}
\caption{Pulse-phase residuals for \ppks\ using the alternative timing models presented in Table~\ref{tab:pksalt} that do not involve a glitch.  These fits include the year 2000 \chandra\ data point (see Section~\ref{sec:one} for details). Top: A quadratic model fit leaves a sinusoidal-like oscillation in the phase residuals. Bottom: the residuals from a cubic model fit (including the frequency second derivative).
}
\label{fig:pksalt}
\end{figure}

\begin{deluxetable}{lc}
\tighten
\tablewidth{0pt}
\tablecolumns{2}
\tablecaption{Alternative Timing Solutions for \ppks}
\tablehead{
\colhead{Parameter} & \colhead{Value\tablenotemark{a} \hfill}
}
\startdata
\cutinhead{Quadratic Timing Solution (2000--2023)\hfill}
 Epoch of ephemeris (MJD TDB)\tablenotemark{b}             &      55478.00000467               \\
 Span of ephemeris (MJD)                   &      51549--59986                   \\
 Frequency, $f$                      &      2.357763483778(12) s$^{-1}$         \\
 Frequency derivative, $\dot f$  &      $-1.1128(15)\times 10^{-16}$  s$^{-2}$ \\
 Period, $P$                          &      0.4241307522490(21) s        \\
 Period derivative, $\dot P$               &      $2.0018(28)\times 10^{-17}$    \\
 $\chi^2_{\nu}[{\rm DoF}]$                 &       8.55[64]                     \\
\cutinhead{Cubic Timing Solution (2000--2023) \hfill} 
 Epoch of ephemeris (MJD TDB)\tablenotemark{b}              &      55478.00000453           \\
 Span of ephemeris (MJD)                   &      51549--59986             \\
 Frequency, $f$                            &      2.357763483219(32) s$^{-1}$       \\
 Frequency derivative, $\dot f$            &   $-1.1260(17)\times10^{-16}$ s$^{-2}$  \\
 Frequency second derivative, $\ddot f$    &      $4.67(25)\times 10^{-26}$ s$^{-3}$ \\
 Period, $P$                               &      0.4241307523495(58) s       \\
 Period derivative, $\dot P$               &      $2.0255(30)\times 10^{-17}$    \\
 Period second derivative, $\ddot P$       &      $-8.40(45)\times 10^{-27}$ s$^{-1}$   \\
 $\chi^2_{\nu}[{\rm DoF}]$                 &       3.17[63]                     \\[-0.5em]
\cutinhead{Cubic Timing Solution (2002--2023) \hfill} 
 Epoch of ephemeris (MJD TDB)\tablenotemark{b}              &       55478.00000447           \\ 
Span of ephemeris (MJD)                   &      52266--59986                   \\
 Frequency, $f$                            &      2.357763483142(35) s$^{-1}$            \\
 Frequency derivative, $\dot f$  &   $-1.1318(20)\times10^{-16}$ s$^{-2}$  \\
 Frequency second derivative, $\ddot f$    &      $5.51(29)\times 10^{-26}$ s$^{-3}$    \\
 Period, $P$                               &      0.42413075236(63) s           \\
 Period derivative, $\dot P$               &      $2.0359(36)\times 10^{-17}$    \\
 Period second derivative, $\ddot P$       &      $-0.92(53)\times 10^{-26}$ s$^{-1}$   \\
 $\chi^2_{\nu}[{\rm DoF}]$                 &       2.72[62]
\enddata
\tablenotetext{a}{$1\sigma$ uncertainties in the last digits are given in parentheses.}
\tablenotetext{b}{Epoch of the ephemeris corresponds to the minimum of the pulse.}
\label{tab:pksalt}
\end{deluxetable}

The uncertainty on the frequency derivative of the revised postglitch timing solution has grown, perhaps an effect of timing noise. In addition, the postglitch frequency derivative is larger (less negative) than the preglitch value by $\approx9\%$, albeit at the $4\sigma$ level. This change is much larger than the $\Delta\dot f/\dot f\sim10^{-3}$ typically seen across glitches, which also suggests that timing noise may predominate.  Moreover, the sign of the change of $\dot f$ is unusual and has implications that will be discussed in Section~\ref{sec:discussion}.


Weak glitches can be difficult to detect in the presence of timing noise, and timing noise can masquerade as glitches \citep{ant22}. A possible example of such an ambiguity is manifest in the downward curvature of the residuals around the time of the assumed glitch in Figure~\ref{fig:pksglitch}. Following \cite{got20}, we therefore considered alternative timing models that do not require the use of a glitch to characterize the timing behavior of \ppks. First, we show that a fit to the data using a quadratic model only, parameterized by the frequency and first derivative, and including the year 2000 observation, leaves a strong sinusoidal-like residual in the phases, with an amplitude of $\sim 0.1$~cycles (see Figure~\ref{fig:pksalt}). As in \citet{got20}, this results in an unacceptable fit statistic of $\chi^2_{\nu} = 8.549$ for 64 DoF, substantially worse than for the glitch model. The sinusoidal residual with a period similar to the data span suggests that a higher-degree polynomial term should be added to the model. The lower panel of Figure~\ref{fig:pksalt} shows the result of fitting a cubic model that includes a frequency second derivative. This reduces the statistic to $\chi^2_{\nu} = 3.17$ for 63 DoF, still significantly higher than for the glitch model. By excluding the 2000 data point, the fit statistic improves marginally to $\chi^2_{\nu} = 2.72$ for 62 DoF but still suggests larger than expected timing noise, as will be discussed in Section~\ref{sec:discussion}. 

The parameters of the quadratic and cubic fits are given in Table~\ref{tab:pksalt}. It is important that each of these models, as well as the glitch model, have the same cycle count between all adjacent ToAs.  Thus, they are all the same phase-connected solution that we believe is the true one, not an alias.  Finally, we note that the binary orbit model suggested in \cite{got20} remains a viable interpretation, although less likely a priori.

\subsection{\ppsr\ in \psnr}

\begin{figure}
\centerline{\includegraphics[width=0.55\textwidth]{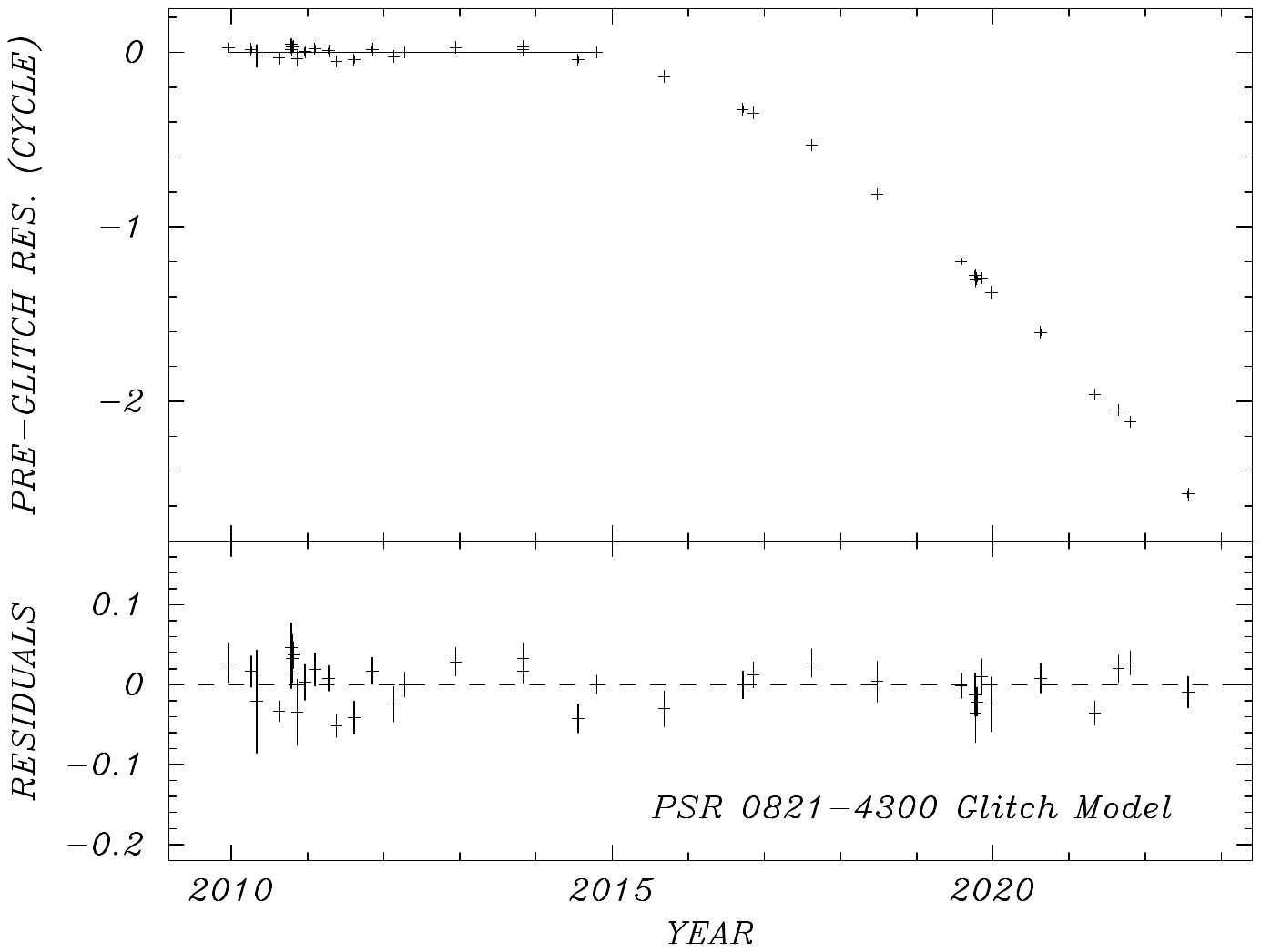}}
\caption{Glitch model for the spin-down of \ppsr. Top: pulse-phase residuals from the preglitch timing solution presented in Table~\ref{tab:pupaglitch}.  The glitch epoch of  2015 December 6 is estimated by matching the frequency of the pre- and postglitch solutions. Bottom: combined residuals from fits to independent timing models for the pre- and postglitch intervals, respectively. The overall $\chi^2_{\nu}$ statistic for the fit is 2.20 for 32 DoF, taking into account the fit parameters for each interval.}
\label{fig:pupaglitch}
\end{figure}

\begin{deluxetable}{lc}
\tighten
\tablewidth{0pt}
\tablecolumns{2}
\tablecaption{Glitch Ephemerides for \ppsr}
\tablehead{
\colhead{Parameter} & \colhead{Value\tablenotemark{a} \hfill} 
}
\startdata
Epoch of position and $\mu$ (MJD)     & 53964.0 \\
R.A. (J2000)                                  & $08^{\rm h}21^{\rm m}57.\!^{\rm s}3653(31)$ \\
Decl. (J2000)                                 & $-43^{\circ}00^{\prime}17.\!^{\prime\prime}074(43)$ \\
R.A. proper motion, $\mu_{\alpha}\,{\rm cos}\,\delta$  & $-54.1\pm 8.3$ mas yr$^{-1}$ \\
Decl. proper motion, $\mu_{\delta}$           & $-28.1\pm 10.5$ mas yr$^{-1}$ \\
Surface dipole magnetic field, $B_s$          & $2.9 \times 10^{10}$ G\\
Spin-down luminosity, $\dot E$                & $1.9 \times 10^{32}$ erg s$^{-1}$ \\
Characteristic age, $\tau_c$                  & 254 Myr \\
\cutinhead{Preglitch Timing Solution (2010--2014) \hfill}
 Epoch of ephemeris (MJD TDB)\tablenotemark{b}              &          56065.00000030           \\
 Span of ephemeris (MJD)                   &      55182--56948                   \\
 Frequency, $f$                       &      8.865291025552(85) s$^{-1}$             \\
 Frequency derivative, $\dot f$  &      $-6.801(44)\times 10^{-16}$ s$^{-2}$   \\
 Period, $P$                            &      0.1127994554401(11) s         \\
 Period derivative, $\dot P$               &      $8.654(56)\times 10^{-18}$     \\
 $\chi^2_{\nu}[{\rm DoF}]$                 &       2.70[19]                     \\
\cutinhead{Postglitch Timing Solution (2015--2022) \hfill}  
 Epoch of ephemeris (MJD TDB)\tablenotemark{b}              &          58530.00000022           \\
Span of ephemeris (MJD)                   &      57272--59787                   \\
 Frequency, $f$                    &      8.865290891814(75) s$^{-1}$            \\
 Frequency derivative, $\dot f$  &      $-6.321(27)\times 10^{-16}$ s$^{-2}$   \\
 Period, $P$                          &      0.11279945714171(95) s       \\
 Period derivative, $\dot P$               &      $8.042(34)\times 10^{-18}$     \\
 $\chi^2_{\nu}[{\rm DoF}]$                 &       1.47[13]                     \\             
 \noalign{\vskip 0.5em}\hline\noalign{\vskip 0.5em}
 Glitch epoch (MJD)\tablenotemark{c}                       &      57362                          \\
 $\Delta f$                          &      $6.26(55)\times 10^{-9}$ s$^{-1}$    \\
 $\Delta f/f_{\rm pred}$                         &      $7.06(62)\times 10^{-10}$ \\
 $\Delta \dot f$                        & $4.80(52)\times 10^{-17}$ s$^{-2}$ \\
 $\Delta \dot f/\dot f$                         &      $-0.0706(76)$
\enddata
\tablenotetext{}{Notes. Position and proper motion from \cite{may20}; Derived parameters ($B_s$, $\dot E$, $\tau_c$) are based on the preglitch timing solution.}
\tablenotetext{a}{$1\sigma$ uncertainties in the last digits are given in parentheses.}  
\tablenotetext{b}{Epoch of the ephemeris corresponds to the maximum of the pulse.}
\tablenotetext{c}{Epoch of the glitch estimated by matching the frequency of the two timing solutions; this assumes a constant postglitch $\dot f$.}
\label{tab:pupaglitch}
\end{deluxetable}

\begin{figure}
\centerline{\includegraphics[width=0.55\textwidth]{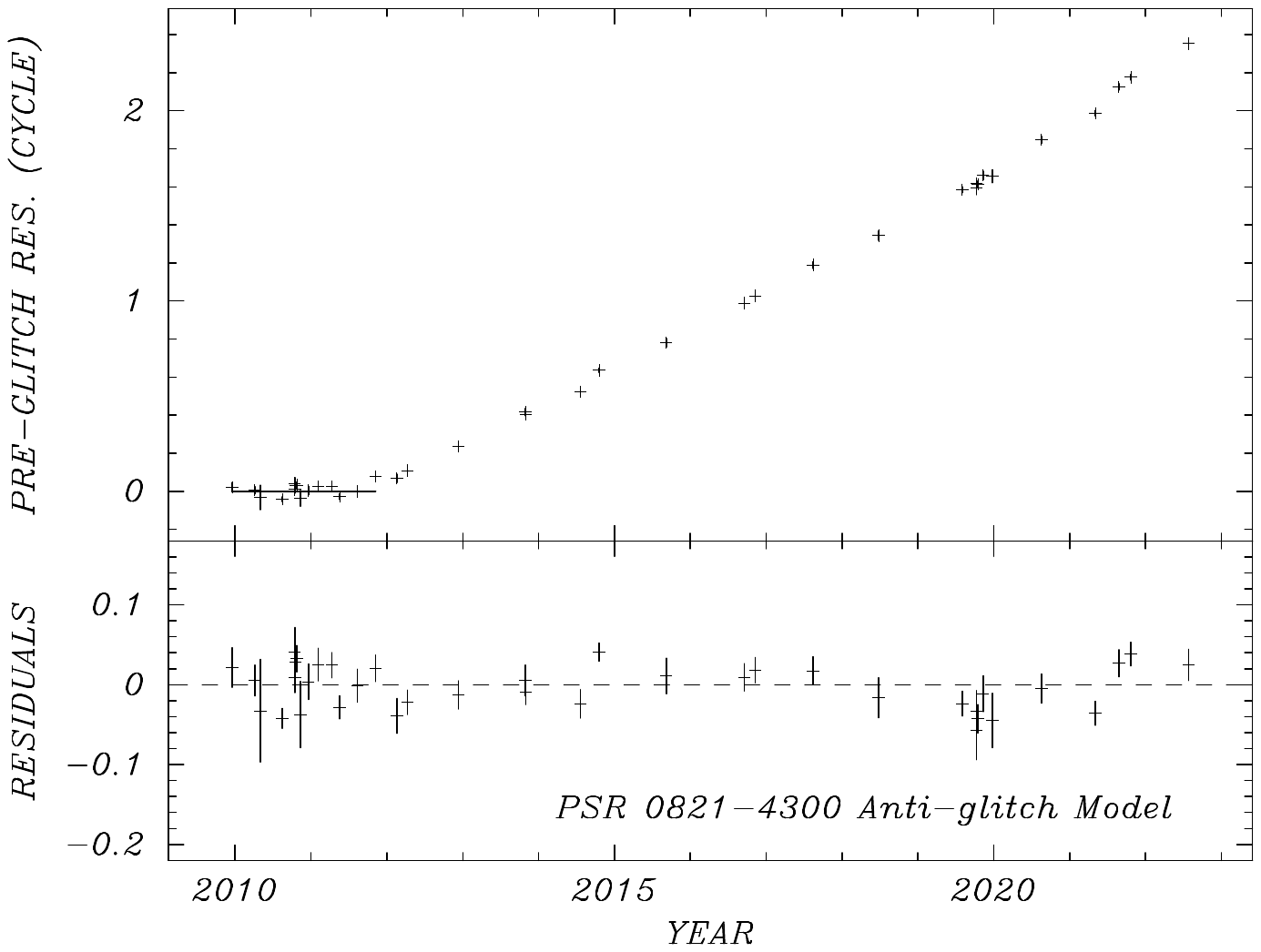}}
\caption{Antiglitch model for the spin-down of \ppsr. Top: pulse-phase residuals from the preglitch timing solution presented in Table~\ref{tab:pupaantiglitch}.  The glitch epoch of 2011 September 14 is estimated by matching the frequency of the pre- and postglitch solutions. Bottom: combined residuals from fits to independent timing models for the pre- and postglitch intervals, respectively. The overall $\chi^2_{\nu}$ statistic for the fit is 2.77 for 32 DoF, taking into account the fit parameters for each interval.}
\label{fig:pupaantiglitch}
\end{figure}

\begin{deluxetable}{lc}
\tighten
\tablewidth{0pt}
\tablecolumns{2}
\tablecaption{Antiglitch Ephemerides for \ppsr}
\tablehead{
\colhead{Parameter} & \colhead{Value\tablenotemark{a}} 
}
\startdata
\cutinhead{Preglitch Timing Solution (2010--2011) \hfill}
 Epoch of ephemeris (MJD TDB)\tablenotemark{b}              &          55484.00000014           \\
 Span of ephemeris (MJD)                   &      55182--55784                   \\
 Frequency, $f$                      &      8.86529106058(42) s$^{-1}$             \\
 Frequency derivative, $\dot f$ &      $-6.14(50)\times 10^{-16}$  s$^{-2}$   \\
 Period, $P$                           &      0.1127994549943(53)  s          \\
 Period derivative, $\dot P$               &      $7.81(64)\times 10^{-18}$      \\
 $\chi^2_{\nu}[{\rm DoF}]$                 &       2.535[11]                 \\
\cutinhead{Postglitch Timing Solution (2012--2022) \hfill}  
 Epoch of ephemeris (MJD TDB)\tablenotemark{b}              &          57830.00000060           \\
 Span of ephemeris (MJD)                   &      55873--59787                   \\
 Frequency, $f$                      &      8.865290929387(33)   s$^{-1}$          \\
 Frequency derivative, $\dot f$ &      $-6.2179(81)\times 10^{-16}$ s$^{-2}$  \\
 Period, $P$                            &      0.11279945666364(42) s          \\
 Period derivative, $\dot P$               &      $7.912(10)\times 10^{-18}$     \\
 $\chi^2_{\nu}[{\rm DoF}]$                 &       2.695[21]                     \\
 \noalign{\vskip 0.5em}\hline\noalign{\vskip 0.5em}
 Glitch epoch (MJD)\tablenotemark{c}                        &      55828                       \\
 $\Delta f$                         &      $-5.4(6)\times 10^{-9}$  s$^{-1}$    \\
 $\Delta f/f_{\rm pred}$                         &      $-6.1(6)\times 10^{-10}$ 
\enddata
\tablenotetext{a}{$1\sigma$ uncertainties in the last digits are given in parentheses.}  
\tablenotetext{b}{Epoch of the ephemeris corresponds to the maximum of the pulse.}
\tablenotetext{c}{Epoch of the glitch estimated by matching the frequency of the two timing solutions; this assumes a constant postglitch $\dot f$.}
\label{tab:pupaantiglitch}
\end{deluxetable}

As listed in the Appendix, the timing observations of \ppsr\ used in this work include the \chandra\ and \xmm\ data sets previously reported in \cite{got13} that span the years 2009--2012, but exclude the two observations from 2001.  In retrospect, it is not possible to span the large time gap from 2001 to 2009 due to the uncertainty in extrapolation of the preglitch ephemeris.  In creating ToAs for \ppsr, we restricted the extracted photon energies to the 1.5--4.5~keV range because of the unique energy dependence of the pulse phase. Below 1.5~keV, the pulse is shifted $\approx180^{\circ}$, effectively canceling the pulsations if the full energy band is used \citep[see ][for details]{got09}.  Photon arrival times were extracted using circular apertures of radius $30^{\prime\prime}$ and $1\farcs2$ for the \xmm\ and \chandra\ datasets, respectively.

Following the results presented in \cite{got13}, we were able to continue monitoring this pulsar and extend its ephemeris but only up to 2014. Subsequently, the yearly monitoring showed increasing departures of the pulse arrival times from the refined ephemeris. Motivated by the similarity to the glitch-like behavior of \ppks, we then initiated a renewed joint \xmm\ and \chandra\ timing campaign on \ppsr\ in 2019-2020 to generate an independent phase-connected timing solution for comparison.

In Figure~\ref{fig:pupaglitch}, we postulate that a glitch in pulse arrival time likely occurred between the 2015 and 2016 observations. Indeed, the monotonic deviations from the preglitch timing solution given in Table~\ref{tab:pupaglitch} imply a change in frequency of magnitude $\Delta f/f_{\rm pred} = 7.1(6)\times 10^{-10}$, a value similar to that found for \ppks.  Including the pre- and postglitch timing solutions, the statistic for the combined data set is $\chi^2_{\nu} = 2.20$ for 32 DoF, taking into account the fit parameters for each interval. 

The sparse coverage around the time of the glitch, between 2014 and 2017, prevents a definitive description of the glitch behavior.  However, formal fits show curvature in the trend of postglitch residuals relative to the preglitch solution, corresponding to a $\approx7\%$ change in frequency derivative, which becomes less negative, similar to the case of \ppks\ but at a higher $9\sigma$ level of significance (see Table~\ref{tab:pupaglitch}).

\begin{figure}
\centerline{\includegraphics[width=0.55\textwidth]{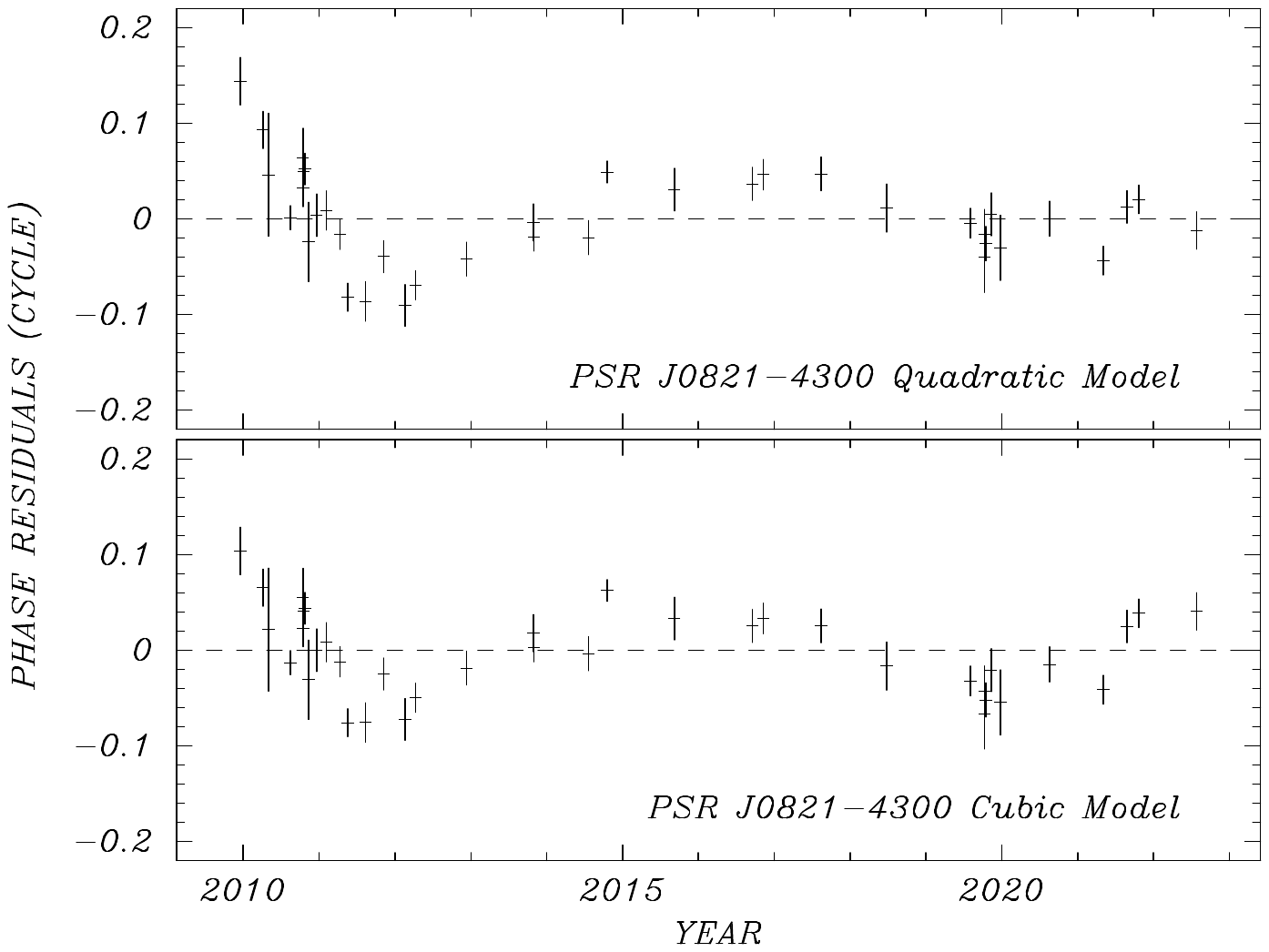}}
\caption{Pulse-phase residuals for \ppsr\ using the alternative timing models presented in Table~\ref{tab:pupaalt} that do not involve a glitch.  
Top: a quadratic model. Bottom: a cubic model (including the frequency second derivative).
}
\label{fig:pupaalt}
\end{figure}

\begin{deluxetable}{lc}
\tighten
\tablewidth{0pt}
\tablecolumns{2}
\tablecaption{Alternative Timing Solutions for \ppsr}
\tablehead{
\colhead{Parameter} & \colhead{Value\tablenotemark{a} \hfill}
}
\startdata
\cutinhead{Quadratic Timing Solution (2010--2022)\hfill}
 Epoch of ephemeris (MJD TDB)\tablenotemark{b}              &          57485.00000096           \\
 Span of ephemeris (MJD)                   &      55182--59787                   \\
 Frequency, $f$                      &      8.865290948152(24) s$^{-1}$        \\
 Frequency derivative, $\dot f$  &      $-6.2723(51)\times 10^{-16}$ s$^{-2}$  \\
 Period, $P$                          &      0.11279945642488(30) s        \\
 Period derivative, $\dot P$               &      $7.9807(65)\times 10^{-18}$    \\
 $\chi^2_{\nu}[{\rm DoF}]$                 &       6.79[35]           \\        
\cutinhead{Cubic Timing Solution (2010--2022) \hfill} 
 Epoch of ephemeris (MJD TDB)\tablenotemark{b}              &          57485.00000097           \\
 Span of ephemeris (MJD)                   &      55182--59787                   \\
 Frequency, $f$              &      8.865290947727(71)   s$^{-1}$       \\
 Frequency derivative, $\dot f$ &      $-6.2665(52)\times 10^{-16}$  s$^{-2}$ \\
 Frequency second derivative, $\ddot f$    &      $1.00(16)\times 10^{-25}$ s$^{-3}$      \\
 Period, $P$                            &      0.11279945643030(91)  s         \\
 Period derivative, $\dot P$  & $7.9733(66)\times 10^{-18}$   \\
 Period second derivative, $\ddot P$     &      $-1.27(20)\times 10^{-27}$   s$^{-1}$  \\
 $\chi^2_{\nu}[{\rm DoF}]$    &       5.81[34]                   \\ 
\enddata
\tablenotetext{a}{$1\sigma$ uncertainties in the last digits are given in parentheses.}
\tablenotetext{b}{Epoch of the ephemeris corresponds to the maximum of the pulse.}
\label{tab:pupaalt}
\end{deluxetable}

Curiously, a model involving an antiglitch (decrease in frequency) fits nearly as well as the glitch model, with the event occurring at an earlier time, around 2011 September (Figure \ref{fig:pupaantiglitch}). The parameters of this model are given in Table~\ref{tab:pupaantiglitch}, where it can be seen that the magnitude of the antiglitch is similar to that of the glitch model, while the statistic for the combined data set is slightly worse at $\chi^2_{\nu} = 2.77$.
Since the preglitch segment in this model is very short, the initial frequency derivative was only measured to a precision of $\approx8\%$, but it is consistent with the postglitch value. Note that this model has the same set of cycle counts as the glitch model, despite being a different analytic and qualitative description of the same timing data.  

As in the case of \ppks, the smooth curvature of the residuals around the epoch of the assumed glitch in Figure~\ref{fig:pupaglitch} leads us to consider alternative timing models without the use of a glitch to characterize the timing of \ppsr. A basic quadratic model leaves phase residuals with $\sim 0.1$~cycle amplitude, similar in magnitude to those of \ppks, but not clearly sinusoidal in nature (Figure~\ref{fig:pupaalt}). The fit statistic is $\chi^2_{\nu} = 6.79$ for 35 DoF, again substantially worse than for the glitch model and similar to that found for \ppks. The addition of a second frequency derivative term for a cubic model has little effect on the shape of the residual curve, and only reduces
$\chi^2_{\nu}$ to 5.80 for 34 DoF. Parameters from the quadratic and cubic fits are given in Table~\ref{tab:pupaalt}. It can be seen that the residual curve in Figure~\ref{fig:pupaalt} could also be interpreted as an early antiglitch or a later glitch. Again, none of these are distinct timing solutions (aliases); they all have the same set of cycle counts.

\subsection{\kpsr\ in \ksnr}

\kpsr\ was the first CCO to have its spin-down rate measured \citep{hal10}, but there followed a 12~yr hiatus in its monitoring. As listed in the Appendix, we obtained a new set of observations of \kpsr\ from a joint \chandra\ and \xmm\ investigation designed to restart a phase-connected timing solution in 2021--2023. Photon arrival times were extracted in the 1--5~keV energy range using a $12^{\prime\prime}$ or $18^{\prime\prime}$ radius aperture with \xmm,\ and a $1\farcs8$ radius aperture with \chandra. 
 
\begin{figure}
\centerline{\includegraphics[width=0.47\textwidth]{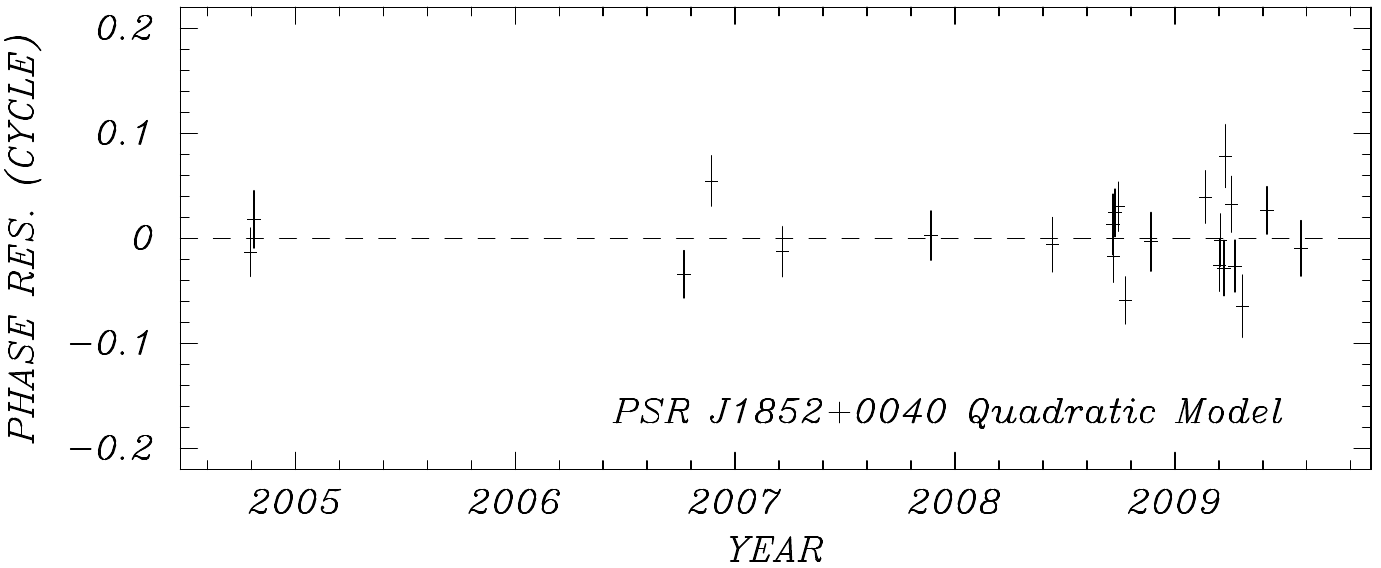}}
\centerline{\includegraphics[width=0.47\textwidth]{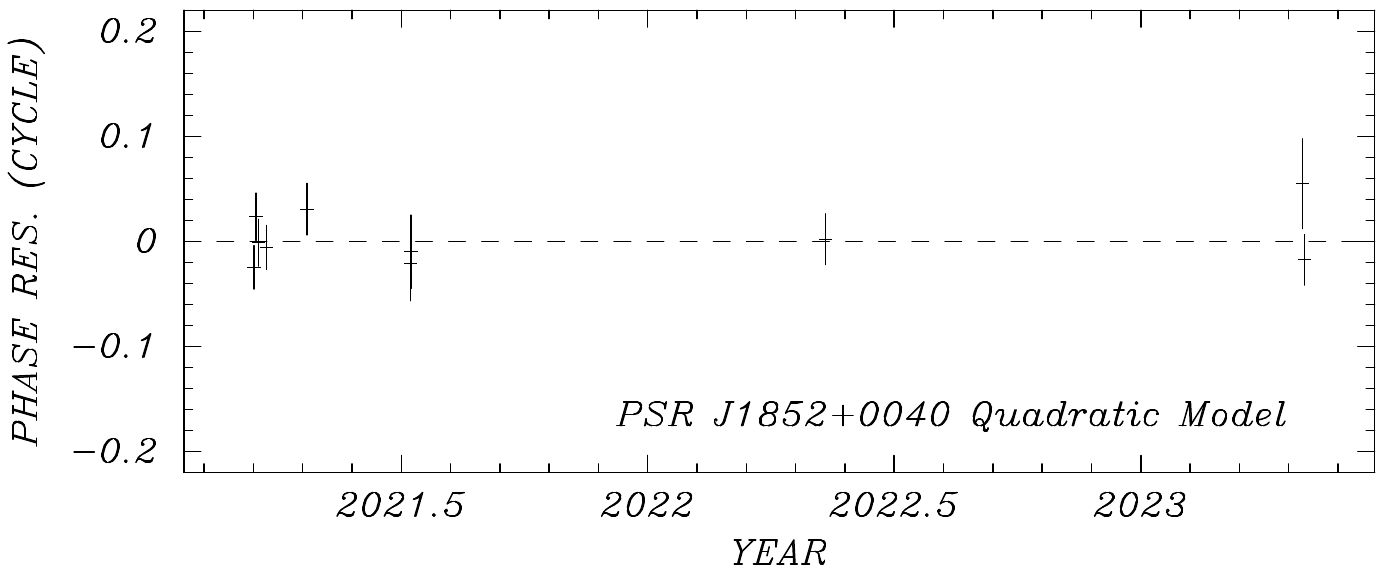}}
\caption{Pulse-phase residuals for \kpsr\ from the quadratic spin-down model presented in Table~\ref{tab:kes79}, 
from two independent timing solutions corresponding to the intervals 2004--2009 (top) and  2021--2023 (bottom). 
}
\label{fig:kes79}
\end{figure}

Table~\ref{tab:kes79} presents the new results from the 10 observations obtained in 2021--2023, 
along with results from the 2004--2009 era \citep{hal10}. As shown in Figure~\ref{fig:kes79}, we are able to obtain independent timing solutions for \kpsr\ both before and after the 12~yr gap. However, the long gap does not allow the cycle count to be maintained, given the uncertainty on the frequency derivative of the earlier solution. 

During the 2021--2023 epoch, the frequency derivative is poorly constrained in the fit, as it contributes less than 2 cycles, based on the prior measured value in 2004--2009. The ephemeris for the earlier era predicts the frequency in 2021--2023 to within $\Delta f = 4.05\times10^{-9}$~Hz, comparable to the extrapolated uncertainty on this parameter ($\sigma_f = 4.03 \times\ 10^{-9}$~Hz). This is of a similar magnitude as that measured for both the \ppks\ and \ppsr\ glitches.
Thus, we would not be sensitive to a glitch in \kpsr\ having similar magnitude as inferred for the two other CCO pulsars.

\begin{deluxetable}{lc}
\tighten
\tablewidth{0pt}
\tablecolumns{2}
\tablecaption{Timing Solutions for \kpsr}
\tablehead{
\colhead{Parameter} & \colhead{Value\tablenotemark{a}}
}
\startdata
R.A. (J2000)                               & $18^{\rm h}52^{\rm m}38^{\rm s}\!.57$ \\
Decl. (J2000)                              & $+00^{\circ}40^{\prime}19^{\prime\prime}\!.8$ \\
Surface dipole magnetic field, $B_s$       & $3.1 \times 10^{10}$~G \\
Spin-down luminosity, $\dot E$             & $3.0 \times 10^{32}$~erg~s$^{-1}$ \\
Characteristic age, $\tau_c$               & 192~Myr \\
\cutinhead{Quadratic Timing Solution (2004--2009)\hfill}
 Epoch of ephemeris (MJD TDB)\tablenotemark{b}              &          54168.00000093           \\
 Span of ephemeris (MJD)                   &      53296--55041                   \\
 Frequency, $f$                      &      9.53174258208(17)  s$^{-1}$       \\
 Frequency derivative, $\dot f$  &      $-7.882(81)\times 10^{-16}$  s$^{-2}$  \\
 Period, $P$                          &      0.1049126108253(19)  s          \\
 Period derivative, $\dot P$               &      $8.676(90)\times 10^{-18}$     \\
 $\chi^2_{\nu}[{\rm DoF}]$                 &       1.01[20]                     \\
\cutinhead{Quadratic Timing Solution (2021--2023) \hfill} 
 Epoch of ephemeris (MJD TDB)\tablenotemark{b}              &          59677.00000016           \\
 Span of ephemeris (MJD)                   &      59288--60066                   \\
 Frequency, $f$                       &      9.53174220286(34) s$^{-1}$         \\
 Frequency derivative, $\dot f$  &      $-8.62(45)\times 10^{-16}$  s$^{-2}$   \\
 Period, $P$                         &      0.1049126149991(37)  s          \\
 Period derivative, $\dot P$               &      $9.48(50)\times 10^{-18}$      \\
 $\chi^2_{\nu}[{\rm DoF}]$                 &       0.976[7]                   
 \enddata
\tablenotetext{}{Notes. Derived parameters ($B_s$, $\dot E$, $\tau_c$) are based on the 2004--2009 timing solution.}
\tablenotetext{a}{$1\sigma$ uncertainties in the last digits are given in parentheses.}
\tablenotetext{b}{Epoch of the ephemeris corresponds to the minimum of the pulse.}
\label{tab:kes79}
\end{deluxetable}

\section {Discussion}
\label{sec:discussion}

\subsection{Do CCOs Glitch?}

The new timing observations reported here extend our previous ephemeris for \ppks, \citep{got20} and reveal a similar noisy timing behavior in \ppsr.  However, because of the sparse sampling and the relatively imprecise ToAs of thermal X-ray pulses compared with those of radio pulsars, these results do not conclusively prove that glitches occur in CCOs.  A glitch is a priori surprising in a CCO, as discussed in \citet{got20}, because CCOs have $|\dot f| < 10^{-15}$~s$^{-2}$.  The glitch activity parameter for an individual pulsar is defined as $$\dot f_g \equiv {\sum_j \Delta f_j \over T},$$ where the numerator is the sum of the frequency steps $\Delta f_j$ over glitches and $T$ is the total time span of monitoring; therefore, $\dot f_g/|\dot f|$ approximates the long-term fraction of the spin-down that is reversed by glitches.

\citet{bas22} find that $\dot f_g$ correlates best with $\dot f$ (and inversely with $\tau_c$) such that $\dot f_g/|\dot f|=0.018\pm0.003$ in the range $10^{-14}<|\dot f|<10^{-10.5}$ s$^{-2}$.  However, it is rare that any pulsar with $|\dot f| < 10^{-15}$~s$^{-2}$ will glitch.  As summarized in \citet{bas22}, the proportionality of glitch activity to spin-down rate can be understood as reflecting the differential velocity that must build up between superfluid and normal components before stress becomes large enough to trigger the unpinning of superfluid vortices.  The same could apply if unpinning is triggered by a crustquake.  According to \citet{ant22}, PSR B0410+69 is the only pulsar with $\dot f$ as small as $\approx -5\times10^{-16}$~s$^{-2}$ that was observed to glitch.  CCOs, which have a similar or smaller spin-down rate, are statistically not expected to glitch.

Ultimately, there is a limit in which a weak glitch can be either obscured or mimicked by timing noise, and that limit may have been reached in our study.  Another manifestation of this ambiguity is the ability of an early antiglitch to fit the timing of \ppsr\ in our data almost as well as a later glitch does.  We consider the antiglitch fit for \ppsr\ to be unconvincing as a physical model because it rests on a short preglitch span of data.  Although antiglitches are sometimes seen in magnetars and accreting pulsars, they are rare in rotation-powered pulsars (PSR~B0540$-$69: \citealt{tuo24}, PSR~J1522$-$5735: \citealt{pan25}).  Also, \citet{esp24} favor a fit of the same event in PSR~B0540$-$69 as changes in $\dot f$ with no evident step in $f$, i.e., similar to timing noise.  In Section~\ref{sec:noise} we discuss timing noise as an alternative to fits involving glitches.  

With these caveats in mind, the implications of the changes in frequency derivative $\dot f$ required in our glitch fits deserve explicit mention. The newly extended postglitch timing solution for \ppks, by refining its frequency derivative from the previous study, revealed a large relative change of $\approx9\%$ in $\dot f$ across the putative glitch. That is, $\Delta\dot f/\dot f \approx -0.09$, with $\dot f$ postglitch becoming less negative.  Note that the magnitude of $\Delta\dot f \sim 10^{-17}$~s$^{-2}$ is itself not unprecedented across glitches (e.g., Figure~11 of \citealt{ant22}), but its relative change and sign are unusual.  These are different from the more common glitches in which $\Delta\dot f/\dot f \approx 10^{-3}$ with $\dot f$ becoming more negative in 80\% of the cases \citep{ant22}.  Although the change in $\dot f$ in \ppks\ is only a $4\sigma$ result, more definitive evidence of the same effect is found in the case of \ppsr; its $\dot f$ became $7\%$ less negative after the hypothesized glitch, the change being of $9\sigma$ significance.

If the surface dipole magnetic field strength $B_s$ is naively estimated as $\propto f^{-3/2} \dot f^{1/2}$, the fitted changes in $\dot f$ imply a decrease of $B_s$ by a few percent within a decade, a much shorter timescale than expected from the ages of the host SNRs.  The sign of the change is also in conflict with a leading theory for the absence of old CCOs \citep{ho11}, which postulates that their weak external dipole field $B_s$ was caused by prompt burial of a normal $B$ field by supernova fallback debris, while reemergence (increase of $B_s$) by ohmic diffusion should follow on timescales of thousands of years \citep{mus95}, allowing the CCOs to join the population of canonical pulsars in the $P-\dot P$ diagram.

It should be noted that comparable or larger variations in $\Delta\dot f/\dot f$ have been discovered from several pulsars in recent years, but they are all associated with changes in magnetospheric emission.  These phenomena include (1) intermittent pulsars in which the radio pulsations turn off for periods of days to years \citep{kra06,cam12,lyn17} while $|\dot f|$ decreases by $\ge 50\%$, (2) quasiperiodic changes in radio pulse profile that are associated with a few percent changes in $\dot f$ \citep{lyn10}, and (3) the radio-quiet PSR J2021+4026 that has correlated states of $\gamma$-ray flux and $\dot f$, differing by $\approx20\%$ and $\approx10\%$, respectively \citep{all13,fio24}.  However, none of these effects can easily be applied to CCOs, with their steady surface thermal X-ray emission and pulse profiles, small spin-down power, and no apparent magnetospheric generation of particles and winds. More likely, therefore, the years-long variation of $\dot f$ in CCOs has its origin in internal properties that are in common with other young but more energetic pulsars, such as high temperature, strong intrinsic $B$ field and superfluid behavior; in some combination, these properties may cause glitches \citep{ho15} or timing noise. However, with the discovery of weak radio pulsations from \ppks\ \citep{zha25}, especially if they are intermittent, magnetospheric effects may be reconsidered as a driver of variations in $\dot f$.

\subsection{Do CCOs have Unusual Timing Noise?}
\label{sec:noise}

Fits to the spin-down of \ppks\ and \ppsr\ using only $f$ and $\dot f$ without invoking glitches leave systematic trends in their phase residuals that may be regarded as timing noise.  Here we attempt to evaluate how this compares to timing noise in canonical pulsars with similar spin-down properties.  Several metrics of timing noise were reviewed by \citet{nam19}.  The most useful here is $\sigma_{\rm TN,2}$ from \citet{sha10}, in which $$\sigma_{\rm TN,2}^2 = \sigma_{\rm R,2}^2 - \sigma_{\rm W}^2.\eqno(1)$$ Here $\sigma_{\rm R,2}$ is the rms of the residuals measured from a quadratic fit, and $\sigma_{\rm W}$ is the typical uncertainty of a ToA.  The subscripts R and W indicate a red or white noise process.

An advantage of this method is that it explicitly acknowledges the red noise character of the timing noise by including in the analysis the total monitoring time span $T$.  \citet{sha10} found for hundreds of canonical pulsars that the mean value of $\sigma_{\rm TN,2}$ scales with the spin parameters as $$\bar\sigma_{\rm TN,2} = C_2\,f^{\alpha}\,|\dot f|^{\beta}\,T^{\gamma} \ \ \mu{\rm s},\eqno(2)$$ where $C_2 = 41.7, \alpha=-0.9, \beta=1.0$, and $\gamma=1.9$. Recognizing the large scatter in $\sigma_{\rm TN,2}$ around a given mean $\bar\sigma_{\rm TN,2}$, \citet{sha10} modeled its distribution as log-normal and found that the standard deviation of ln($\sigma_{\rm TN,2}$) = 1.6.

Applying this method to the quadratic fit of \ppks\ (Figure~\ref{fig:pksalt}), the rms timing residual is $\sigma_{\rm R,2}=25.69$~ms, and the average uncertainty of a ToA is $\sigma_{\rm W}=10.95$~ms; therefore, $\sigma_{\rm TN,2}=23.24$~ms.  In comparison, $\bar\sigma_{\rm TN,2}\approx148\,\mu$s from Equation (2) for the timing parameters of \ppks.  The residuals of \ppks\ therefore exceed the population average by a factor of $\approx157$, which is 5 standard deviations of the distribution of ln($\sigma_{\rm TN,2}$).  This anomaly is confirmed relative to the data in the various panels in Figure~6 of \citealt{nam19}, which graph $\sigma_{\rm TN,2}$ as a function of several spin-down parameters for a sample of 91 pulsars.  Another way to describe this result is that the timing noise of \ppks\ is comparable to that of pulsars with 2 orders of magnitude greater values of $|\dot f|$ or $B_s$.

The same analysis of the residuals of the quadratic fit of \ppsr\ in Figure~\ref{fig:pupaalt} gives $\sigma_{\rm R,2}=5.20$~ms, $\sigma_{\rm W}=2.44$~ms and $\sigma_{\rm TN,2}=4.59$~ms.  In comparison, $\bar\sigma_{\rm TN,2}\approx80\,\mu$s from Equation (2).  The residuals of \ppsr\ therefore exceed the expected average by a more modest factor of $\approx57$, or 4 standard deviations in the natural log.  In the case of \kpsr, it is not surprising that the monitoring periods were too short to display residuals of the same magnitude as in \ppsr.  Nevertheless, the significant excess of timing residuals in two out of the three CCO pulsars is sufficient to argue that CCOs as a class have unexpectedly large timing noise relative to pulsars with similar spin-down properties.

Another characterization of timing noise simply uses the frequency second derivative (equivalent to the braking index) added to the fit as an approximate measure of the residuals.  An example is the cubic fit of \ppks\ in Figure~\ref{fig:pksalt}, which yields a braking index of $n = f\ddot f \dot f^{-2}=8.6\times10^6$.  In comparison, pulsars with positive $\ddot f$ generally have $n<2\times10^5$ (Figure 5 of \citealt{nam19}).  In the case of \ppsr, it is interesting that fitting a second derivative of $\ddot f=1.0\times10^{-25}$~s$^{-3}$ barely improves the fit in Figure~\ref{fig:pupaalt}, while the equivalent braking index is $n=2.3\times10^{5}$. For either pulsar, this method indicates large timing noise, but the meaning is not precise as it does not take into account the time span of the observations or the nature of the residuals remaining after the cubic fit.  For completeness, we recall that an earlier parameterization of timing noise by \citet{arz94} used the frequency second derivative measured over a time span of $T=10^8$~s to define $$\Delta_8 = {\rm log} \left({1\over6f}\,|\ddot f|\,T^3\right).$$ However, we do not have enough precise ToAs within any time span of $10^8$~s to measure a significant $\ddot f$.

Finally, we note that the fitted second derivatives representing timing noise are large enough, given the long time spans of the data, to slowly produce the large fractional changes in $\dot f$ that, in context of the glitch model, were constrained to be discrete events.  Neither interpretation of the changes in $\dot f$ appears to be clearly preferable, although timing noise is perhaps simpler.
 
\section {Conclusions}
\label{sec:conclusions}

Continued timing of the known CCO pulsars establishes \ppsr\ as the second one with timing noise or glitches, quantitatively similar to the behavior of \ppks\ established previously.  We have taken care to ensure that all of the models applied to the timing residuals of a particular pulsar have the same set of cycle counts between observations.  Thus, they are the same timing solution fitted with different analytic formulas and not aliases.  Even so, distinguishing glitches from timing noise in these CCOs is difficult because of the infrequent sampling and lower precision of the ToAs that can be obtained from these thermal X-ray pulsations compared to those of typical radio pulsars. Nevertheless, whichever description is adopted, the magnitude of the timing irregularities is greater than in rotation-powered pulsars with a comparably small spin-down rate and a dipole magnetic field strength as small as that of CCOs.

A quantitative result of extended monitoring is that the spin-down rate $\dot f$ in \ppks\ and \ppsr\ appears to decrease by $7\%-9\%$ on a timescale of a decade.  This is probably not a monotonic trend that represents a decrease in the dipole $B$ field but more likely a representative short-term fluctuation.  In the absence of evidence for magnetospheric activity and intermittent mode changes such as are seen in certain variable rotation-powered pulsars, the origin of the timing variations in CCOs should be sought in the internal structure of these young NSs, which may have superfluid dynamics affected by high temperatures and $B$ fields much stronger than their external dipole component.  Accretion torque noise from a residual fallback disk is even a possible contributor.  Noisy timing behavior is an important feature of CCOs as a class, and it may inform theories of glitches and timing noise in the general pulsar population.

\begin{acknowledgments}
This paper employs a list of Chandra datasets, obtained by the Chandra X-ray Observatory, contained in the Chandra Data Collection ~\dataset[DOI: 10.25574/cdc.523]{https://doi.org/10.25574/cdc.523}. Support for this work was provided by NASA through XMM grants 80NSSC18K0452, 80NSSC19K0866, 80NSSC20K1310, 80NSSC21K0819, 80NSSC22K0867; \nicer\ grants 80NSSC19K1461, 80NSSC21K0106, 80NSSC21K1884, 80NSSC22K1300; and Chandra Awards SAO GO7-18063X, SAO GO0-21059X, SAO GO1-22065X issued by the Chandra X-ray Observatory Center, which is operated by the Smithsonian Astrophysical Observatory for and on behalf of NASA under contract NAS8-03060. This investigation is based partly on observations obtained with \xmm, an ESA science mission with instruments and contributions directly funded by ESA Member States and NASA. This research has made use of data and/or software provided by the High Energy Astrophysics Science Archive Research Center (HEASARC), which is a service of the Astrophysics Science Division at NASA/GSFC.
\end{acknowledgments}

\appendix

The logs of observations of \ppks, \ppsr, and \kpsr, are presented in Tables~\ref{obslogpks},~\ref{obslogpupa} and \ref{tab:obslogkes79}, respectively.

\startlongtable
\begin{deluxetable}{llclc}
\tablewidth{0pt}
\tablecolumns{5}
\tablecaption{Log of X-ray Timing Observations of \ppks}
\tablehead{
\colhead{Mission}  & \colhead{Instrument/Mode}   & \colhead{ObsID\tablenotemark{a}}  & \colhead{Date} & \colhead {Exp\tablenotemark{b}} \\
\colhead{} & \colhead{} & \colhead{} & \colhead{(UT)} & \colhead{(ks)} 
}
\startdata
Chandra & ACIS-S3/CC        & 751            & 2000 Jan 6   & 32.4  \\
XMM    & EPIC-pn/sw        & 0113050501     & 2001 Dec 23   & 27.0  \\  
Chandra & ACIS-S/CC         & 2799           & 2002 Jan 5   & 30.3  \\
XMM    & EPIC-pn/sw        & 0155960301     & 2002 Aug 4    & 128.0 \\
XMM     & EPIC-pn/sw        & 0155960501     & 2002 Aug 6    & 129.0 \\
Chandra & ACIS-S/CC         & 3915           & 2003 Jun 10   & 155.1 \\
Chandra & ACIS-S/CC         & 4398           & 2003 Jun 18   & 114.7 \\
XMM    & EPIC-pn/sw        & 0304531501     & 2005 Jun 22   & 15.1  \\
XMM     & EPIC-pn/sw        & 0304531601     & 2005 Jul 5   & 18.2  \\
XMM   & EPIC-pn/sw        & 0304531701     & 2005 Jul 10   & 20.5  \\
XMM     & EPIC-pn/sw        & 0304531801     & 2005 Jul 11   & 63.4  \\
XMM     & EPIC-pn/sw        & 0304531901     & 2005 Jul 12   & 14.5  \\
XMM    & EPIC-pn/sw        & 0304532001     & 2005 Jul 17   & 16.5  \\ 
XMM    & EPIC-pn/sw        & 0304532101     & 2005 Jul 31   & 17.7  \\
XMM     & EPIC-pn/sw        & 0552810301     & 2008 Jul 2   & 31.4  \\
XMM    & EPIC-pn/sw        & 0552810401     & 2008 Dec 22   & 30.4  \\
Chandra & ACIS-S3/CC        &      14199     & 2011 Nov 25    & 31.0  \\
Chandra & ACIS-S3/CC        &      14202     & 2012 Apr 10    & 33.0  \\
XMM     & EPIC-pn/sw        & 0679590101     & 2012 Jun 22   & 26.5  \\
XMM   & EPIC-pn/sw        & 0679590201     & 2012 Jun 24   & 22.3  \\
XMM     & EPIC-pn/sw        & 0679590301     & 2012 Jun 28   & 24.9  \\
XMM    & EPIC-pn/sw        & 0679590401     & 2012 Jul 2   & 24.5  \\
XMM    & EPIC-pn/sw        & 0679590501     & 2012 Jul 18   & 27.3  \\
XMM     & EPIC-pn/sw        & 0679590601     & 2012 Aug 11    & 27.3  \\
Chandra & ACIS-S3/CC        &      14200     & 2012 Dec 1   & 31.1  \\
Chandra & ACIS-S3/CC        &      14203     & 2013 May 19    & 33.0  \\
Chandra & ACIS-S3/CC        &      14201     & 2013 Dec 4   & 33.0  \\
Chandra & ACIS-S3/CC        &      14204     & 2014 Jun 20   & 33.0  \\
XMM    & EPIC-pn/sw        &0780000201      & 2016 Jul 28   & 32.5  \\
XMM     & EPIC-pn/sw        &0800960201      & 2017 Jun 22   & 33.3  \\
XMM     & EPIC-pn/sw        &0800960301      & 2017 Jun 23   & 20.7  \\
XMM    & EPIC-pn/sw        &0800960401      & 2017 Jun 24   & 22.6  \\
XMM    & EPIC-pn/sw        &0800960501      & 2017 Jul 3   & 23.5  \\
NICER & XTI  &1020270102      & 2017 Jul 24   & 6.1   \\  
NICER & XTI  &1020270106      & 2017 Jul 28   & 14.3  \\  
NICER & XTI  &1020270110      & 2017 Aug 1    & 12.1  \\  
XMM   & EPIC-pn/sw        &0800960601      & 2017 Aug 10    & 19.8  \\
Chandra & ACIS-S3/CC        &   19612        & 2017 Oct 10   & 32.9  \\
NICER & XTI  &1020270130      & 2017 Nov 15    & 20.6  \\  
XMM    & EPIC-pn/sw        &0800960701      & 2017 Dec 24   & 19.8  \\
XMM    & EPIC-pn/sw        &0821940201      & 2018 Jun 22   & 33.2  \\ 
Chandra & ACIS-S3/CC        & 19613          & 2018 Aug 27    & 66.3  \\  
NICER & XTI &1020270153-58   & 2018 Nov 30    & 7.6   \\  
XMM  & EPIC-pn/sw &0821940301 & 2018 Dec 28   & 26.6  \\
NICER & XTI &2506010101-02   & 2019 Apr 4   & 22.3  \\  
XMM & EPIC-pn/sw &0842280301 & 2019 Jul 9   & 30.8  \\ 
NICER & XTI &2506010201-02   & 2019 Jul 19   & 21.4  \\ NICER & XTI &2506010205-13   & 2019 Jul 26   & 6.7   \\ 
\cutinhead{ New Observations -- This Work}
NICER & XTI &2506010214-24   & 2019 Aug 29    & 10.6 \\ 
NICER & XTI &2506010301-03   & 2019 Oct 31   & 15.6 \\ 
NICER & XTI &2506010304-409  & 2019 Nov 19    &  9.8 \\ 
NICER & XTI &2506010415-20   & 2020 Feb 16    & 19.0 \\ 
NICER & XTI &3550010101-03   & 2020 Mar 30    & 17.8 \\ 
NICER & XTI &3550010201-03   & 2020 Jun 12   & 18.6 \\ NICER & XTI &3550010401-03   & 2020 Dec 23   & 17.7 \\ 
NICER & XTI &3550010501-02   & 2021 Feb 27    & 17.3 \\ 
NICER & XTI &4614010101-03   & 2021 Apr 4   & 16.3 \\ 
NICER & XTI &4614010201-04   & 2021 May 28    & 14.5 \\ 
NICER & XTI &4614010301-04   & 2021 Jun 29   & 12.9 \\ 
NICER & XTI &4614010501-02   & 2021 Nov 19    &  8.2 \\ 
NICER & XTI &4614010601-02   & 2022 Jan 1    &  8.5 \\ 
NICER & XTI &4614010701-02   & 2022 Feb 4   & 17.4 \\ 
NICER & XTI &4614010801-03   & 2022 Apr 19   & 17.7 \\ 
NICER & XTI &4614010901-05   & 2022 Jun 16   & 21.1 \\ 
NICER & XTI &4614011001-02   & 2022 Sep  1   &  8.0 \\ 
NICER & XTI &4614011201-03   & 2023 Feb 11   & 15.8
\enddata
\tablenotetext{a}{The \nicer\ datasets are denoted by the ObsID and date of the first of the concatenated set of short adjacent observations.}
\tablenotetext{b}{Exposure times for the \xmm\ EPIC-pn do not reflect the 29\% deadtime in SmallWindow (sw) mode.}
\label{obslogpks}
\end{deluxetable}

\begin{deluxetable}{llclc}
\tablewidth{0pt}
\tablecolumns{5}
\tablecaption{Log of X-ray Timing Observations of \ppsr}
\tablehead{
\colhead{Mission}  & \colhead{Instrument/Mode}   & \colhead{ObsID}   & \colhead{Date} & \colhead{Exp\tablenotemark{a}} \\
\colhead{} & \colhead{} & \colhead{} & \colhead{(UT)} & \colhead{(ks)} 
}
\startdata
XMM &EPIC-pn/sw& 0606280101 & 2009      Dec 17, 18& 85 \\ 
XMM &EPIC-pn/sw& 0606280201 & 2010      Apr 5 & 42      \\ 
XMM &EPIC-pn/sw& 0650220201 & 2010      May 2 &28      \\ 
Chandra &ACIS-S3/CC &12108  & 2010      Aug 16 &34      \\ 
XMM &EPIC-pn/sw &0650220901 & 2010      Oct 15 &24     \\ 
XMM &EPIC-pn/sw &0650221001 & 2010      Oct 15 &24     \\ 
XMM &EPIC-pn/sw &0650221101 & 2010      Oct 19 &27     \\ 
XMM &EPIC-pn/sw &0650221201 & 2010      Oct 25 &25     \\ 
XMM &EPIC-pn/sw &0650221301 & 2010      Nov 12 &24     \\ 
XMM &EPIC-pn/sw &0650221401 & 2010      Dec 20 &27     \\ 
Chandra &ACIS-S3/CC &12109  & 2011      Feb 4  &33      \\ 
XMM &EPIC-pn/sw &0650221501 & 2011      Apr 12 &30     \\ 
XMM &EPIC-pn/sw &0657600101 & 2011      May 18 &37     \\ 
Chandra &ACIS-S3/CC &12541  & 2011      Aug 11 &33      \\ 
XMM &EPIC-pn/sw &0657600201 & 2011      Nov 8  &37      \\ 
Chandra &ACIS-S3/CC &12542, 14395 & 2012 Feb 18, 19 & 33  \\ 
XMM &EPIC-pn/sw &0657600301 & 2012      Apr 10 &35  \\ 
\cutinhead{ New Observations -- This Work}
Chandra & ACIS-S3/CC & 14798 & 2012 Dec 11 &    31 \\ 
XMM & EPIC-pn/sw & 0722640301 & 2013 Oct 29 &   45  \\ 
XMM & EPIC-pn/sw & 0722640401 & 2013 Oct 31 &   42  \\ 
Chandra & ACIS-S3/CC & 16254 & 2014 Jul 21 &     33 \\ 
XMM & EPIC-pn/sw & 0742040201 & 2014 Oct 18 &    46 \\ 
Chandra & ACIS-S3/CC & 16255 & 2015 Sep 7 &      33 \\ 
Chandra & ACIS-S3/CC & 16256 & 2016 Sep 18 &    34 \\ 
XMM & EPIC-pn/sw & 0781870101 & 2016 Nov 8 &    91 \\ 
Chandra & ACIS-S3/CC & 19609 & 2017 Aug 14 &    12 \\ 
Chandra & ACIS-S3/CC & 19610,21110 & 2018 Jun 26, 28 &    34 \\ 
Chandra & ACIS-S3/CC & 19611 & 2019 Aug 2 &     33 \\ 
XMM & EPIC-pn/sw & 0853220201 & 2019 Oct 9 &     23 \\ 
XMM & EPIC-pn/sw & 0853220301 & 2019 Oct 11 &    26 \\ 
XMM & EPIC-pn/sw & 0853220401 & 2019 Oct 16 &    34 \\ 
XMM & EPIC-pn/sw & 0853220501 & 2019 Nov 10 &    26 \\ 
XMM & EPIC-pn/sw & 0853220601 & 2019 Dec 26 &    23 \\ 
Chandra  & ACIS-S3/CC & 22674 & 2020 Aug 17 & 45\\ 
XMM & EPIC-pn/sw & 0882950101  & 2021 May 3  & 37\\ 
Chandra & ACIS-S3/CC & 22675, 25984 & 2021 Aug 25, 27 & 48 \\ 
XMM & EPIC-pn/sw &0882950201 & 2021 Oct 22 &  35 \\ 
Chandra & ACIS-S3/CC & 22676, 26477 & 2022 Jul 27, 28 & 52
\enddata
\tablenotetext{a}{Exposure times for the \xmm\ EPIC-pn do not reflect the 29\% deadtime in SmallWindow (sw) mode.}
\label{obslogpupa}
\end{deluxetable}

\begin{deluxetable}{llclc}
\tablewidth{0pt}
\tablecolumns{5}
\tablecaption{Log of X-ray Timing Observations of \kpsr}
\tablehead{
\colhead{Mission}  & \colhead{Instrument/Mode}   & \colhead{ObsID}   & \colhead{Date} & \colhead{Exp\tablenotemark{a}} \\
\colhead{} & \colhead{} & \colhead{} & \colhead{(UT)} & \colhead{(ks)} 
}
\startdata
XMM & EPIC-pn/sw  & 0204970201   & 2004 Oct 18 & 30.6 \\
XMM & EPIC-pn/sw  & 0204970301   & 2004 Oct 23 & 30.5 \\
XMM & EPIC-pn/sw  & 0400390201   & 2006 Oct 8 & 29.7 \\
\chandra\ & ACIS-S3/CC  & 6676   & 2006 Nov 23 & 32.2 \\
XMM & EPIC-pn/sw  & 0400390301   & 2007 Mar 20 & 30.5 \\
\chandra\ & ACIS-S3/CC  & 9101   & 2007 Nov 12 & 33.1 \\
\chandra\ & ACIS-S3/CC  & 9102   & 2008 Jun 16 & 31.2 \\
XMM & EPIC-pn/sw  & 0550670201   & 2008 Sep 19 & 21.2 \\
XMM & EPIC-pn/sw  & 0550670301   & 2008 Sep 21 & 31.0 \\
XMM & EPIC-pn/sw  & 0550670401   & 2008 Sep 23 & 34.8 \\
XMM & EPIC-pn/sw  & 0550670501   & 2008 Sep 29 & 33.0 \\
XMM & EPIC-pn/sw  & 0550670601   & 2008 Oct 10 & 36.0 \\
\chandra\ & ACIS-S3/CC  & 9823  & 2008 Nov 21 & 30.1 \\
\chandra\ & ACIS-S3/CC  & 9824   & 2009 Feb 20 & 29.6 \\
XMM & EPIC-pn/sw  & 0550671001   & 2009 Mar 16 & 27.0 \\
XMM & EPIC-pn/sw  & 0550670901   & 2009 Mar 17 & 26.0 \\
XMM & EPIC-pn/sw  & 0550671201   & 2009 Mar 23 & 27.3 \\
XMM & EPIC-pn/sw  & 0550671101   & 2009 Mar 25 & 19.9 \\
XMM & EPIC-pn/sw  & 0550671301   & 2009 Apr 4 & 26.0 \\
XMM & EPIC-pn/sw  & 0550671901   & 2009 Apr 10 & 30.5 \\
XMM & EPIC-pn/sw  & 0550671801   & 2009 Apr 22 & 28.0 \\
\chandra\ & ACIS-S3/CC  & 10128  & 2009 Jun 2 & 33.2 \\
\chandra\ & ACIS-S3/CC  & 10129  & 2009 Jul 29 & 32.2 \\
\cutinhead{ New Observations -- This Work}
XMM  & EPIC-pn/sw & 0872790101 & 2021 Mar 15  &  36.4 \\
XMM  & EPIC-pn/sw & 0872790201 & 2021 Mar 16  &  31.5 \\
XMM  & EPIC-pn/sw & 0872790301 & 2021 Mar 18  &  29.6 \\
XMM  & EPIC-pn/sw & 0872790401 & 2021 Mar 24  &  29.6 \\
XMM & EPIC-pn/sw & 0872790501 & 2021 Apr 23  &  29.6 \\
\chandra\ & ACIS-S3/CC & 23866 & 2021 Jul 8  &  21.0 \\
\chandra\ & ACIS-S3/CC & 25085 & 2021 Jul 9  &  21.0 \\
\chandra\ & ACIS-S3/CC & 23867 & 2022 May 12  &  43.0 \\
\chandra\ & ACIS-S3/CC & 23868 & 2023 Apr 30  &  10.1 \\
\chandra\ & ACIS-S3/CC & 27823 & 2023 May 2  &  31.8
\enddata
\tablenotetext{a}{Exposure times for the \xmm\ EPIC-pn do not reflect the 29\% deadtime in SmallWindow (sw) mode.}
\label{tab:obslogkes79}
\end{deluxetable}

\newpage
\bibliography{cco_timing_2023}
\bibliographystyle{aasjournal}

\end{document}